\documentclass[twocolumn]{aastex631}

\shorttitle{Rotation Period of 3I/ATLAS from Jet PA and Photometry}
\shortauthors{Scarmato, Loeb}

\begin{document}

\title{Rotation Period of 3I/ATLAS After Perihelion from Jet Position Angle Wobble and Photometric Variability}

\author{Toni Scarmato}
\affiliation{Toni Scarmato's Astronomical Observatory, San Costantino di Briatico, Calabria, 89817 Italy}

\author{Abraham Loeb}
\affiliation{Astronomy Department, Harvard University, 60 Garden Street, Cambridge MA 02138 USA}

\begin{abstract}
We determine the post-perihelion rotation period of 3I/ATLAS  using two independent diagnostics: the temporal modulation of the position angle (PA) of a persistent jet-like feature, and a time-series photometric light curve in the Gr ($\approx R$) band.
For the jet morphology, we measure the PA at multiple epochs by applying the Larson-Sekanina Rotational Gradient filter to Hubble Space Telescope images between November 20, 2025 and December 27, 2025, and model the phase-folded PA curve with weighted least-squares Fourier series up to two harmonics while scanning trial periods $P$ to identify minima in $\chi^2(P)$.
For the photometry, we adopt the best-fit period from an independent 30-minute binned analysis (from a 0.25 meter telescope MPC L92) based on a refined $\chi^2(P)$ profile for a sinusoidal model with nightly offsets.
We find that the jet-PA modulation gives a period $P_{\rm jet}=7.20\pm0.05$~h (adopting a conservative uncertainty dominated by sparse sampling and systematics), while the photometry yields $P_{\rm phot}=7.136\pm0.001$~h (formal 1$\sigma$ uncertainty).
Although the periods differ slightly, the offset is plausibly attributable to non-Gaussian systematics and aliasing.
The combined data supports a post-perihelion rotation period of 7.1 h associated with precession of the jet structure around the rotation axis by 20°. The rotation axis is aligned with the sunward direction to within 20°.
\end{abstract}

\keywords{comets: general --- Interstellar Interloper: individual (3I/ATLAS) --- methods: data analysis --- techniques: photometric}

\section{Introduction}
The rotation of comets nuclei \citep{delaFuente2005} affects the temporal modulation of their out-gassing, the formation and persistence of jets, and the evolution of coma morphology \citep{HaqueLopez2025};
\citep{Eubanks2025DirectSpacecraft}.
Rotation periods are commonly inferred either from time-resolved photometry (reflecting changes in projected cross section and rotating activity patterns) or from morphological tracers such as repeating jet or spiral features whose apparent geometry is modulated by nucleus rotation and viewing geometry \citep[e.g.,][]{Samarasinha2004,FarnhamSchleicher2005}.
Both diagnostics can be affected by aliasing (single- vs.\ double peak modulation), evolving activity, and non-Gaussian systematics arising from coma contamination, variable seeing, and calibration residuals.
A robust period inference uses photometry and morphology as complementary constraints \citep{SerraRicartLicandroAlarcon2025}.

Time-series period searches are often performed on unevenly sampled data.
The Lomb--Scargle formalism \citep{Lomb1976} and its generalizations provide standard approaches for identifying periodic signals and evaluating their significance under irregular sampling \citep{Lomb1976,Scargle1982,ZechmeisterKurster2009}.
When the observable is non-sinusoidal, truncated Fourier-series (harmonic) representations provide a compact model that remains solvable by linear least squares at fixed trial period \citep[e.g.,][]{NumericalRecipes2007}.
Here we apply harmonic modeling to a phase-folded jet-PA time series of 3I/ATLAS between November 30, 2025 and December 27, 2025, and compare it with an independent photometric constraint with time series observation between December 9, 2025 and December 22, 2025,  to determine its post-perihelion rotation period \citep{Santana2025}.

\section{Data}
\subsection{Jet PA measurements}
We measured the position angle (PA) of a persistent jet-like feature \citep{KetoLoeb2026}; \citep{KetoLoeb2025};\citep{Loeb2025RNAAS335}; in processed images from HST images of 3I/ATLAS, taken at multiple epochs between November 30, 2025 and December 27, 2025 \footnote{https://mast.stsci.edu}. 
Table~\ref{tab:pa} lists the adopted PA measurements and uncertainties.
Because the measurements are clustered within a small number of nights and are subject to morphology-dependent systematics, we adopt a conservative uncertainty of $\sigma_{\rm PA}=3^\circ$ per measurement.

\subsection{Photometry}
We also consider an independent time-series photometric data set in the Gr ($\approx R$) band from 0.25 m Newton Telescope, MPC L92.
The published analysis used 30-minute binning to mitigate short-timescale scatter, resulting in 14 binned points.
The photometric period reported for that analysis is considered here as an independent constraint and is kept distinct from the jet-derived period.

\begin{deluxetable}{lcc}
\tablecaption{Jet position angle measurements.\label{tab:pa}}
\tablehead{\colhead{UT date} & \colhead{PA (deg)}}
\startdata
2025 11 30.80903 & 290 \\
2025 11 30.86389 & 288 \\
2025 11 30.86875 & 288 \\
2025 11 30.87431 & 285 \\
2025 11 30.87911 & 283 \\
2025 11 30.88403 & 283 \\
2025 12 4.65068 & 259  \\
2025 12 4.65554 & 262 \\
2025 12 4.66110 & 265 \\
2025 12 4.66666 & 268 \\
2025 12 12.88889 & 281 \\
2025 12 12.89306 & 278 \\
2025 12 12.89721 & 277 \\
2025 12 12.90139 & 275 \\
2025 12 12.90556 & 273 \\
2025 12 12.90971 & 270 \\
2025 12 27.66528 & 263 \\
2025 12 27.66943 & 260 \\
2025 12 27.67431 & 258 \\
2025 12 27.67993 & 256 \\
2025 12 27.68193 & 255 \\
\enddata
\tablecomments{UT dates are expressed as YYYY~MM~DD.ddddd. The uncertainty is conservatively taken as 3°.}
\end{deluxetable}

\begin{deluxetable}{lc}
\tablecaption{Summary of the independent photometric period analysis (30-minute bins).\label{tab:phot}}
\tablehead{\colhead{Quantity} & \colhead{Value}}
\startdata
Band / binning & Gr ($R$ band) / 30 min \\
N nights / N bins & 4 / 14 \\
Best period $P_{\rm phot}$ & 7.136 h \\
Formal $1\sigma$ on $P_{\rm phot}$ & 0.001 h \\
Semi-amplitude $A$ & 0.311 mag \\
Peak-to-peak $2A$ & 0.622 mag \\
Jitter $\sigma_{\rm jit}$ & 0.089 mag \\
\enddata
\end{deluxetable}

\begin{deluxetable*}{lcccc}
\tablecaption{Binned photometry used for the period analysis (30-minute bins). \label{tab:phot_points}}
\tablehead{
\colhead{UT mid-time} & \colhead{JD} & \colhead{Phase at $P_{\rm phot}$} & \colhead{$m^\prime$ (mag)} & \colhead{$\sigma_{m}$ (mag)}
}
\startdata
2025-12-09 01:21:57 & 2461018.5569094 & 0.000 & 14.465 & 0.006 \\
2025-12-09 01:51:38 & 2461018.5775207 & 0.069 & 14.234 & 0.005 \\
2025-12-09 02:18:39 & 2461018.5962829 & 0.132 & 14.291 & 0.005 \\
2025-12-09 02:52:31 & 2461018.6198051 & 0.212 & 14.422 & 0.006 \\
2025-12-09 03:06:45 & 2461018.6296840 & 0.245 & 14.441 & 0.034 \\
2025-12-15 02:13:03 & 2461024.5923982 & 0.299 & 13.531 & 0.003 \\
2025-12-15 02:47:25 & 2461024.6162669 & 0.379 & 13.172 & 0.003 \\
2025-12-15 02:59:15 & 2461024.6244780 & 0.407 & 13.617 & 0.022 \\
2025-12-18 02:53:27 & 2461027.6204478 & 0.483 & 13.814 & 0.004 \\
2025-12-18 03:23:04 & 2461027.6410228 & 0.552 & 13.990 & 0.004 \\
2025-12-18 03:49:18 & 2461027.6592406 & 0.613 & 13.898 & 0.005 \\
2025-12-22 22:21:21 & 2461032.4314939 & 0.664 & 13.315 & 0.004 \\
2025-12-22 22:53:44 & 2461032.4539862 & 0.739 & 13.887 & 0.005 \\
2025-12-22 23:07:13 & 2461032.4633438 & 0.771 & 14.030 & 0.019 \\
\enddata
\tablecomments{$m^\prime$ denotes magnitudes corrected to a common photometric scale using the median zeropoint across all frames (see Section~\ref{sec:photometry}). Phases are computed for $P_{\rm phot}=7.136$~h with the earliest binned point as phase zero.}
\end{deluxetable*}

\section{Methods}
\subsection{Jet-PA harmonic model and period scan}
For a trial period $P$, we define the rotational phase
\begin{equation}
\phi_i(P) = \mathrm{frac}\!\left(\frac{t_i-t_0}{P}\right),
\end{equation}
where $t_i$ are observation times (in hours), $t_0$ is an arbitrary reference epoch, and $\mathrm{frac}$ denotes the fractional part.
We model the phase dependence of PA with a truncated Fourier series up to harmonic order $K$:
\begin{equation}
{\rm PA}(\phi) = C + \sum_{k=1}^{K}\left[a_k\sin(2\pi k\phi) + b_k\cos(2\pi k\phi)\right].
\label{eq:pa_model}
\end{equation}
At fixed $P$, all coefficients are obtained by weighted least squares using weights $w_i=1/\sigma_{{\rm PA},i}^2$.
We scan $P$ on a grid and record $\chi^2(P)$ for $K=1$ and $K=2$ models.
To assess whether the second harmonic is warranted, we compare the improvement in fit relative to the increase in model complexity see Figure~\ref{fig:pa_resid}.

\subsection{Photometric period }\label{sec:photometry}
The photometric analysis fit a 2-harmonic sinusoid with nightly offsets to 30-minute binned magnitudes using weighted least squares, and determined the best period from a refined $\chi^2(P)$ profile.
We have determined the reported photometric period and formal uncertainty:
\begin{equation}
P_{\rm phot} = 7.136 \pm 0.001~{\rm h}\quad (1\sigma).
\end{equation}\\

\section{Results}
\subsection{Jet-derived period}
A broad period scan identifies a coherent solution near $P\approx7.2$~h when modeling the PA modulation with $K=2$ harmonics.
The phase-folded curve at $P=7.20$~h is shown in Figure~\ref{fig:pa_phase}, together with the best-fit $K=2$ model.
The $K=2$ model captures the non-sinusoidal shape implied by the data and reduces structured residuals relative to $K=1$ see (Figure~\ref{fig:pa_resid}).
Given the sparse sampling and morphology-dependent systematics, we adopt a conservative jet period
\begin{equation}
P_{\rm jet} = 7.20 \pm 0.05~{\rm h},
\end{equation}
where the uncertainty is intended to encompass plausible systematics beyond formal statistical errors.

\subsection{Photometric constraint}
Figure~\ref{fig:phot_phase} shows the photometry phase-folded at the adopted period $P_{\rm phot}$ (phase 0--1).
The light curve exhibits a coherent single peak modulation consistent with a first-harmonic model.
Because the photometric solution is derived from a separate analysis and may be influenced by daily aliasing and additional ``jitter'' beyond signal to noise ratio formal (SNR) uncertainties, it is best interpreted as an independent constraint on the rotation period rather than a definitive substitute for the morphology-based period \cite{JewittLuu2025}.


\subsection{Photometric period analysis and phase-folded light curve}

We searched for periodic modulation in the $R$-band photometry by fitting a
single-harmonic sinusoidal model, allowing for independent nightly zero-point
offsets to absorb night-to-night systematics (e.g., transparency, coma
variations, calibration drifts). Each measurement provides the mid-exposure
time $t_i$ (JD) and the measured magnitude $m_i$.\\

\subsubsection{Zero-point normalization}
When individual frame zero-points $ZP_i$ were available, we homogenized the
photometry to a common reference zero-point $ZP_{\rm ref}$ (taken as the median
over the dataset) as
\begin{equation}
m_{{\rm corr},i} = m_i + \left(ZP_{\rm ref} - ZP_i\right).
\label{eq:zp_correction}
\end{equation}

\subsubsection{Photometric uncertainties and weights}
Formal magnitude uncertainties were estimated from the reported signal-to-noise ratio ${\rm SNR}_i$ as
\begin{equation}
\sigma_i = \frac{1.0857}{{\rm SNR}_i},
\label{eq:sigma_snr}
\end{equation}
and used as inverse-variance weights $w_i = \sigma_i^{-2}$ in a weighted least
squares (WLS) fit.\\

\subsubsection{Within-night binning}
To reduce short-timescale scatter, data were optionally binned within each
night using a time bin $\Delta t$ (e.g., 5, 10, or 30 minutes). For a bin
containing measurements $j$, the weighted binned time and magnitude were
computed as\\

\begin{equation}
t_{\rm bin} = \frac{\sum_j w_j t_j}{\sum_j w_j}\\
\end{equation}
\begin{equation}m_{\rm bin} = \frac{\sum_j w_j m_{{\rm corr},j}}{\sum_j w_j}\\
\end{equation}
\begin{equation}
\sigma_{\rm bin} = \left(\sum_j w_j\right)^{-1/2}
\end{equation}

\
\subsubsection{Single-harmonic model with nightly offsets}
At a trial period $P$ (frequency $f=1/P$; angular frequency
$\omega = 2\pi/P$), we fitted the binned (or unbinned) magnitudes with
\begin{equation}
m_i = C + O_{k(i)} + a \cos(\omega t_i) + b \sin(\omega t_i),
\label{eq:model_1harm}
\end{equation}
where $C$ is a global constant, $O_{k(i)}$ is the additive offset for the night
$k$ to which point $i$ belongs, and $a,b$ are the harmonic coefficients. To fix
the reference level we set
\begin{equation}
O_{1} = 0,
\label{eq:offset_constraint}
\end{equation}
so that all other offsets are measured relative to the first night. The
semi-amplitude of the modulation is
\begin{equation}
A = \sqrt{a^2 + b^2},
\qquad
\Delta m_{\rm p2p} = 2A.
\label{eq:amplitude}
\end{equation}

\subsubsection{Objective function and model selection}
For each trial period $P$, the best-fitting parameters were obtained by WLS
minimization of
\begin{equation}
\chi^2(P) = \sum_i \left(\frac{m_i - m_{\rm model}(t_i;P)}{\sigma_i}\right)^2,
\label{eq:chi2}
\end{equation}
with ${\rm dof} = N - k$ degrees of freedom, where $N$ is the number of points
and $k$ the number of fitted parameters. We also report the Bayesian
Information Criterion
\begin{equation}
{\rm BIC}(P) = \chi^2(P) + k \ln N.
\label{eq:bic}
\end{equation}

\subsubsection{Period search and oversampling}
We searched periods in the range $P_{\min} \le P \le P_{\max}$ by scanning a
uniform frequency grid in cycles per day. For a total time baseline
$T_{\rm span} = t_{\max} - t_{\min}$, the frequency step was set to
\begin{equation}
\Delta f = \frac{1}{{\rm OS}\,T_{\rm span}},
\label{eq:dfreq}
\end{equation}
where ${\rm OS}$ is an oversampling factor. The best period $P_{\rm best}$ was
taken as the minimum of $\chi^2(P)$; we refined the solution with a denser
local grid around the global minimum. A formal $1\sigma$ uncertainty on $P$ was
estimated from the $\Delta\chi^2=1$ criterion:
\begin{equation}
\chi^2(P) = \chi^2_{\min} + 1.
\label{eq:dchi2}
\end{equation}

\subsubsection{Additional scatter (jitter)}
Because SNR uncertainties account primarily for statistical noise, we
estimated an additional ``jitter'' term $\sigma_{\rm jit}$ such that
\begin{equation}
\sigma_{{\rm eff},i}^2 = \sigma_i^2 + \sigma_{\rm jit}^2,
\label{eq:sigma_eff}
\end{equation}
and $\sigma_{\rm jit}$ was chosen to agree approximately with the number of degrees of freedom (dof),
\begin{equation}
\sum_i \left(\frac{r_i}{\sigma_{{\rm eff},i}}\right)^2 \simeq {\rm dof},
\label{eq:jitter_condition}
\end{equation}
where $r_i$ are the residuals of the best-fitting model.

\subsubsection{Phase folding (0--1 cycles) and sinusoidal representation}
For visualization we removed the fitted nightly offsets and folded the data on
$P_{\rm best}$. The offset-corrected magnitudes are
\begin{equation}
\Delta m_i = m_i - \left(C + O_{k(i)}\right),
\label{eq:delta_mag}
\end{equation}
and the rotational phase in the interval $[0,1)$ is defined as
\begin{equation}
\phi_i = {\rm frac}\!\left(\frac{t_i - t_0}{P_{\rm best}}\right),
\qquad 0 \le \phi_i < 1,
\label{eq:phase_def}
\end{equation}
where $t_0$ is an arbitrary reference epoch (e.g., the first observation) and
${\rm frac}(x)=x-\lfloor x \rfloor$ denotes the fractional part. The
corresponding best-fitting sinusoid (in $\Delta m$) is
\begin{equation}
\Delta m_{\rm model}(\phi) = a \cos(2\pi \phi) + b \sin(2\pi \phi).
\label{eq:folded_model}
\end{equation}

\begin{figure*}
\centering
\includegraphics[width=0.95\textwidth]{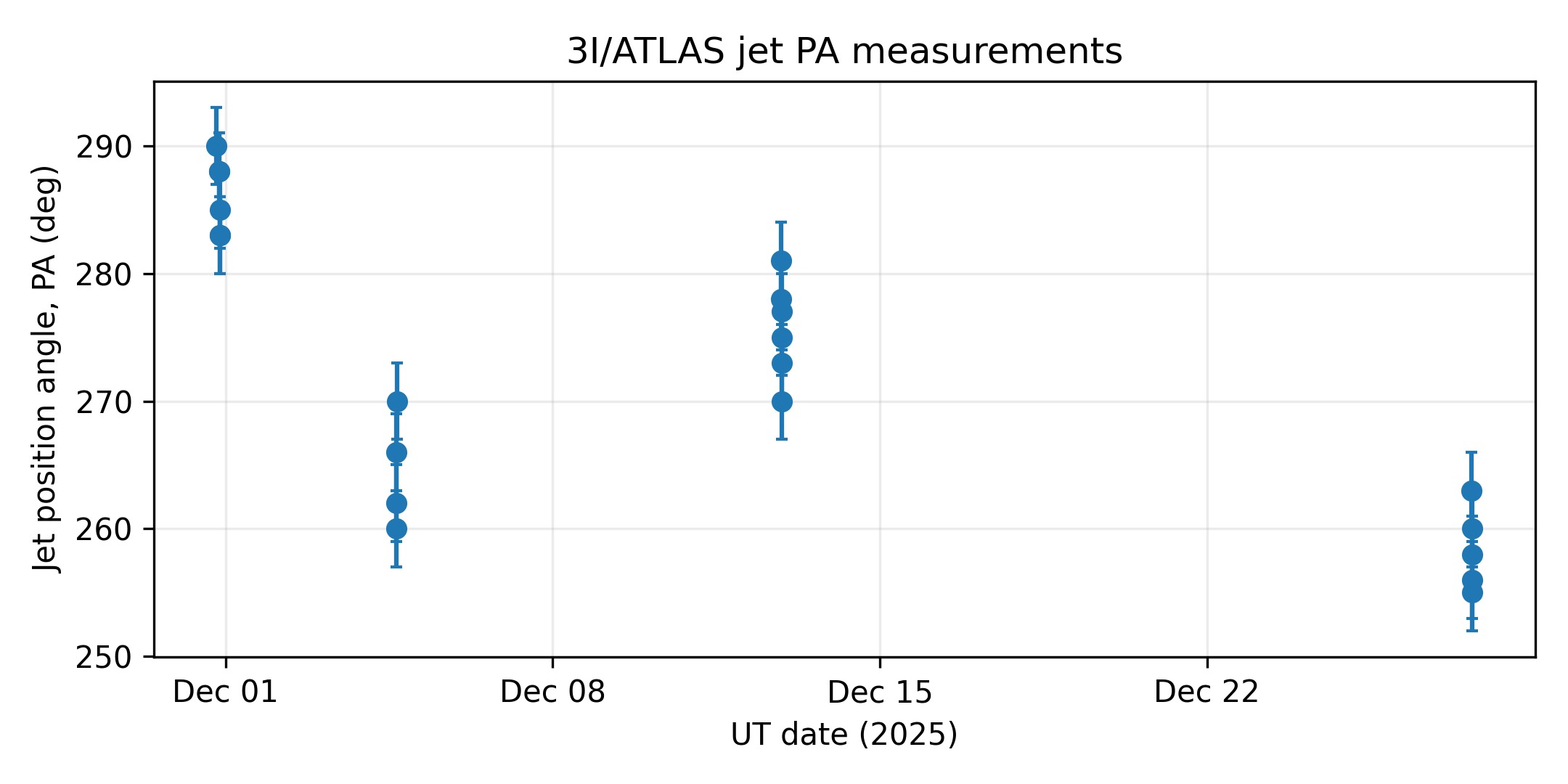}
\caption{Jet position angle (PA) measurements versus time. Error bars show the adopted $\sigma_{\rm PA}=3^\circ$.}
\label{fig:pa_time}
\end{figure*}

\begin{figure*}
\centering
\includegraphics[width=0.95\textwidth]{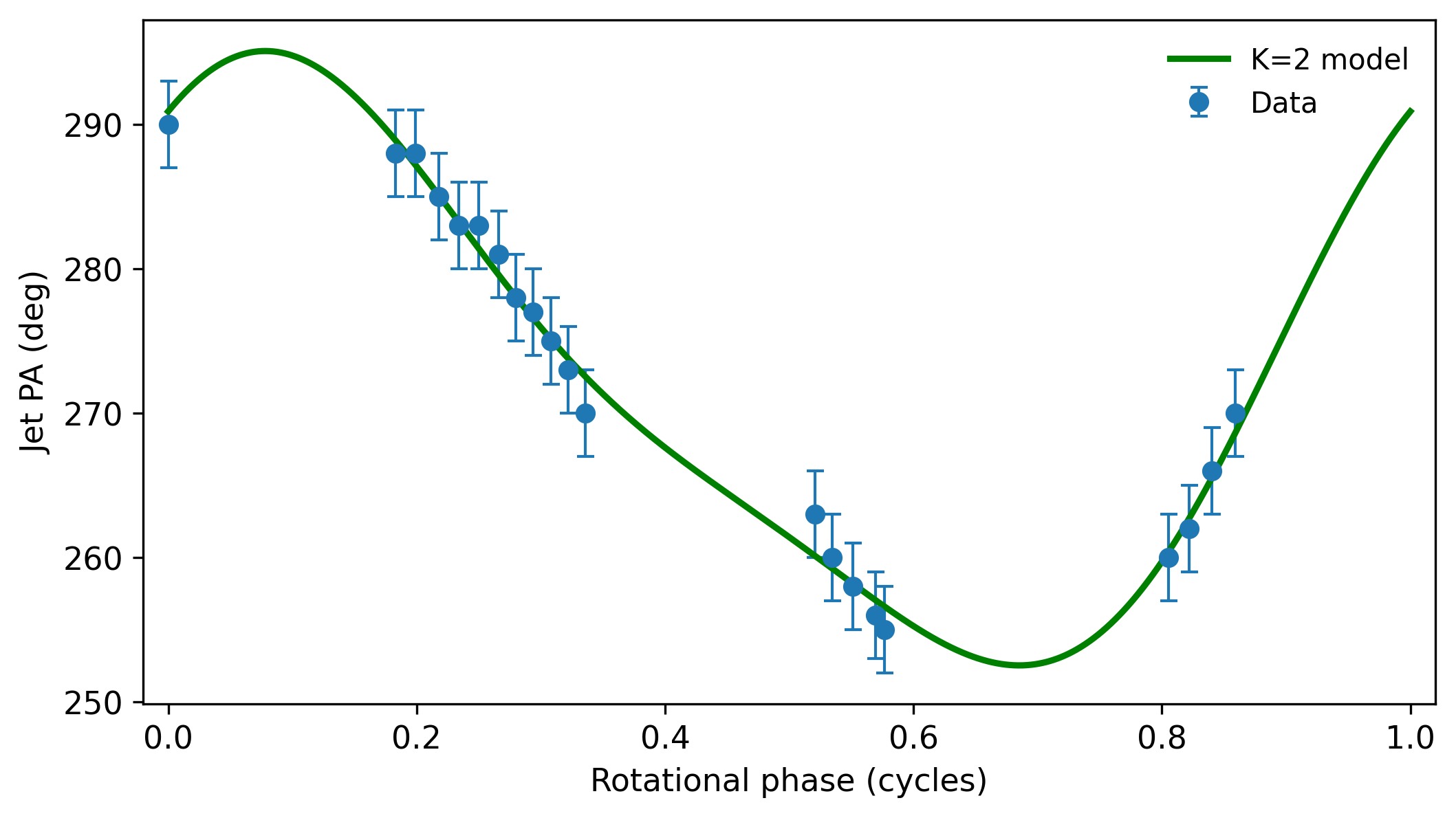}
\caption{Phase-folded jet PA measurements at $P=7.20$~h (phase 0--1). The best-fit $K=2$ model.}
\label{fig:pa_phase}
\end{figure*}

\begin{figure*}
\centering
\includegraphics[width=0.95\textwidth]{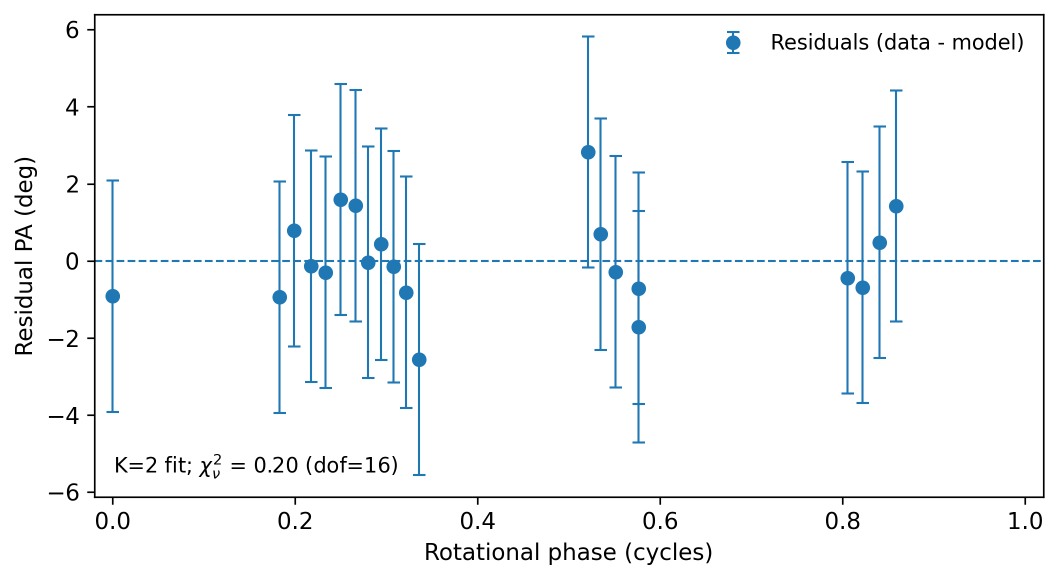}
\caption{Residuals of the jet-PA fits at $P=7.20$~h for $K=2$. The reduced residual structure for $K=2$ supports including a second harmonic.}
\label{fig:pa_resid}
\end{figure*}

\begin{figure*}
\centering
\includegraphics[width=0.95\textwidth]{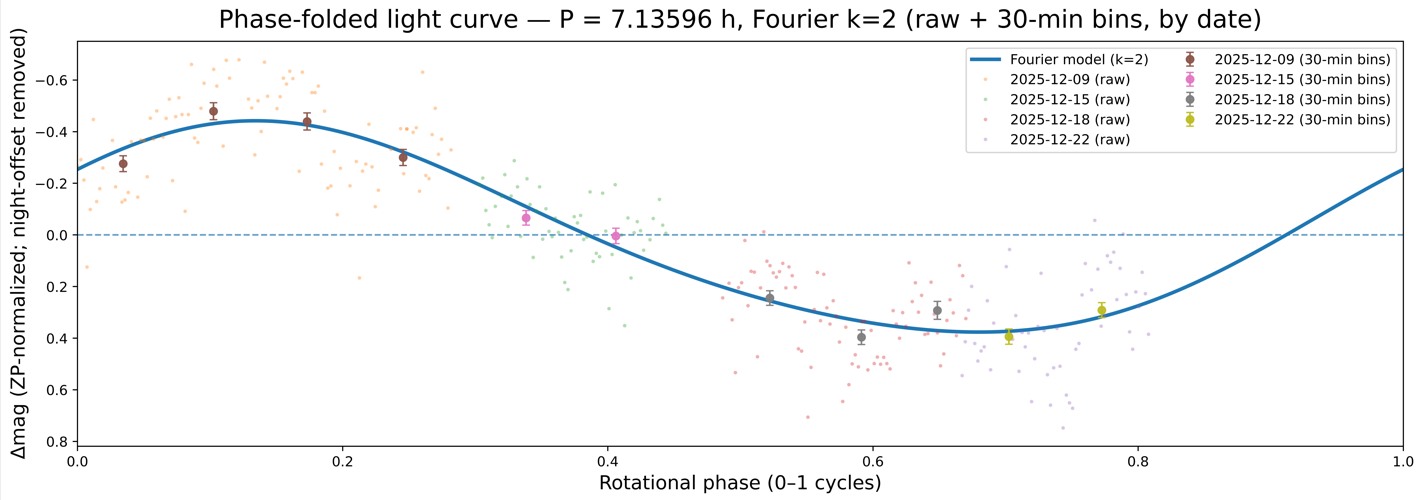}
\caption{Photometry folded at $P_{\rm phot}=7.136$~h (phase 0--1, 30-minute bins). Points are coloured by observing night. The curve is the best-fit  2 harmonic model from the photometric analysis.}
\label{fig:phot_phase}
\end{figure*}

\begin{figure*}
\centering
\includegraphics[width=0.95\textwidth]{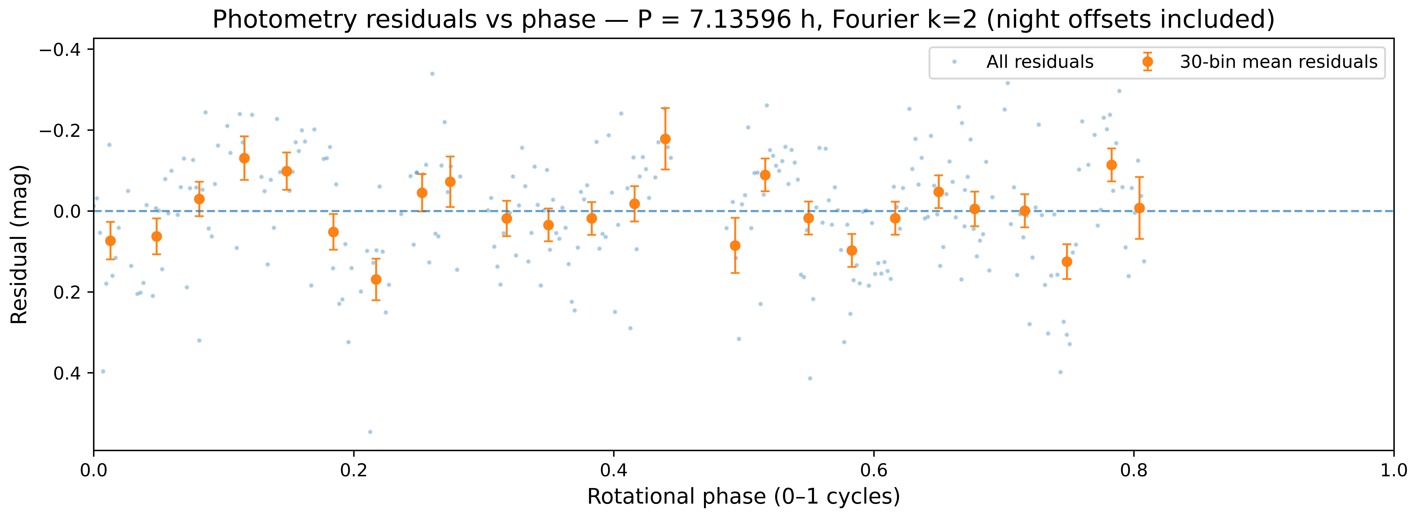}
\caption{ Residual photometry (phase 0-1, 30-minute bins) best-fit  2 harmonic model from the photometric analysis.}
\label{fig:residual_phase}
\end{figure*}

\section{Discussion}
The two diagnostics yield periods that are close but not identical: $P_{\rm jet}\simeq7.20$~h versus $P_{\rm phot}=7.136$~h.
A small offset of this magnitude is not unexpected given: (i) the sparse morphological sampling, (ii) the sensitivity of jet PA to viewing geometry and evolving activity, and (iii) the aliasing and excess scatter in photometry due to coma variability and calibration systematics.
The agreement at the few-percent level nonetheless supports a nucleus rotation timescale 7.1 h. Our inferred value for the PA rotation period after perihelion of $7.20\pm 0.05$ h is consistent with value measured for the periodic PA variations before perihelion, $7.74 \pm 0.35$ h \citep{SerraRicartLicandroAlarcon2025} but is shorter by a factor of $\sim 2.3$ than the rotation period of $\sim 16-17$ h inferred from the photometric variability before perihelion \citep{delaFuente2005}; \citep{Santana2025}. The difference in photometric periodicity might have resulted from a change in the number of jets or active spots - likely induced by the perihelion passage of 3I/ATLAS. The perihelion passage might have doubled the frequency by which the coma flux is modulated by the jets since the rotation axis is nearly aligned with the Sun-3I/ATLAS axis at large heliocentric distance. Whereas only one of the rotation poles was illuminated by the Sun before perihelion, the second rotation pole was illuminated by the Sun after perihelion, potentially triggering two active jets observed from a different viewing angle.
\
\section{Conclusions}
We used two independent approaches to determine the rotation timescale of interstellar interloper 3I/ATLAS \citep{Bolin2025}: harmonic modeling of jet position-angle variability and an independent photometric time series.
We find: (i) a jet-PA wobble period $P_{\rm jet}=7.20\pm0.05$~h  and (ii) photometric period $P_{\rm phot}=7.136\pm0.001$~h.
The two estimates, treated separately, support a rotation period of 7.1 h.
Future dense and contemporaneous monitoring is required to reduce systematic uncertainties, resolve aliasing, and determine whether the modulation is single or double peak.

Our data suggest that the photometric periodicity arises because active regions produce collimated outflows whose projected brightness within a fixed aperture depends on the instantaneous orientation of the jets relative to the observer and the Sun. It was already apparent from the first HST image of 3I/ATLAS \citep{JewittLuu2025} that the nucleus accounts for a sub-percent fraction of the scattered sunlight. As the nucleus rotates, the axis of the dominant outflow direction sweeps around the rotation axis as a result of the precession of the jet, modulating: (i) the column density of dust along the line of sight, (ii) the distribution of dust within the photometric aperture, and (iii) the effective scattering phase function of the dusty coma. Sine the coma provides the dominant contribution to the total flux, even modest changes in jet orientation can produce measurable periodic variations in the integrated magnitude.\footnote{This interpretation was suggested by Loeb (Nov. 30, 2025),\citep{Loeb2025RNAAS178}, in a research note at: \url{https://avi-loeb.medium.com/are-the-jets-from-3i-atlas-pulsed-like-a-heartbeat-fe39dc583d25}}. Importantly, the same underlying rotational state can manifest differently in photometry and in jet position angle (PA). If the jet direction is tied to a fixed active area on a rotating nucleus, the PA can exhibit a periodic modulation, and the integrated coma brightness can vary periodically as the jet alternately points closer to, or farther from, the line of sight. In addition, if the jet direction undergoes precession around the spin axis (e.g., due to a high-latitude source region and changing illumination geometry), the characteristic photometric period corresponds to the jet-orientation cycle rather than directly to a nucleus shape-driven light curve. In our interpretation, the observed periodicity should be regarded as a jet-driven modulation that traces the rotational state through jet orientation, rather than a direct measurement of nucleus rotational light-curve amplitude. 
Since the jet precesses around the rotation axis, we associate its average PA value of 270°+/-3° to be the PA of the rotation axis. Given that the anti-sunward direction is 290°+/-3°, we infer that the rotation axis of 3I/ATLAS is aligned with the Sun-3I/ATLAS axis to within 20 degrees. This alignment needs to be explained, potentially as the result of a torque associated with sublimation of ice on the surface of the nucleus.\\

\begin{acknowledgments}
\textbf{AKNOWLEDGMENTS}
\\ A.L. was supported in part by Harvard's Black Hole Initiative (funded by GBMF and JTF) and the Galileo Project. This work was carried out using Astroart for image processing, Astrometrica for time phorometry,  open-source Python tools for time-series analysis and figure generation, including Astropy, NumPy, SciPy, and Matplotlib.
\end{acknowledgments}

\software{Astropy \citep{Astropy2013,Astropy2018}, NumPy \citep{Harris2020}, SciPy \citep{Virtanen2020}, Matplotlib \citep{Hunter2007}}\\

\bibliographystyle{aasjournal}
\bibliography{references}

\appendix
\section{Deriving the jet position angle from \textit{HST} imaging using Larson--Sekanina processing}
\label{app:hst_ls}

The jet position angle (PA) time series used to determine the rotation period was derived from \textit{Hubble Space Telescope} (\textit{HST}) images by enhancing low-contrast coma structures through a Larson--Sekanina (LS) filter. LS processing is widely used in cometary morphology studies to suppress the quasi-radial coma component and amplify azimuthal intensity gradients associated with jets, fans, and other anisotropic outflow features.

For each \textit{HST} visit, the calibrated frames were registered to a common centroid and (when multiple exposures were available) combined to improve the signal-to-noise ratio while preserving small-scale morphology. The LS operator was applied as a rotational differential filter of the form
\begin{equation}
I_{\rm LS}(r,\theta) = I(r,\theta) - I(r,\theta+\Delta\theta),
\end{equation}
with the rotation increment $\theta$ = 31 ° and $\delta$r = 0.1 chosen to maximize the contrast of the primary jet while avoiding the introduction of spurious multiple-lobe artifacts. In practice, we experimented with a narrow range of $\Delta\theta$ values and adopted the setting that produced the most stable morphology between frames within the same epoch. The resulting enhanced images were inspected to identify the dominant jet ridge line.\\

The jet PA was measured in the conventional sense (east of north) see Figure 6 and Figure 7, by determining the orientation of the enhanced jet axis relative to the photocenter. Measurements were obtained by tracing the jet ridge line over a fixed radial range from the nucleus (chosen to minimize saturation/PSF residuals near the core and to avoid low-S/N regions at large radii) and fitting a straight line to the ridge in the sky plane. The best-fitting orientation was converted to PA, and its uncertainty was estimated by propagating the dispersion among repeated measurements (e.g., varying the radial range and LS parameters within a conservative interval) and by accounting for the finite angular resolution of the enhanced jet feature. The final PA uncertainties adopted in the period fitting are reported in Table~1.

This approach yields a homogeneous PA time series from space-based imaging that is minimally affected by seeing variations and that directly traces the rotational modulation of the active source region driving the observed jet morphology.\\

\section{Ground-based photometry}
\label{app:obs_phot}

The time-series photometry used to derive the light-curve period was acquired at the Toni Scarmato's Astronomical Observatory (observer code: \textit{MPC L92}), using a consistent instrumental setup across the observing campaign. Observations were conducted on multiple nights under photometric or near-photometric conditions, with exposures selected to avoid saturation of field stars and to maintain adequate signal-to-noise ratio on the 3I/ATLAS while minimizing trailing.

All frames were bias/dark corrected and flat-fielded using standard CCD reduction procedures. Astrometric registration was performed to ensure consistent aperture placement across the time series. Photometric calibration used field stars from a standard catalog (\textit{Gaia DR3}), and magnitudes were placed onto a common internal scale through frame-by-frame zeropoint determination \citep{Scarmato2025a}. The formal uncertainty on each measurement was derived from the photometric signal-to-noise ratio, and we additionally accounted for excess scatter arising from transparency variations, background gradients from the coma, and residual systematics (see Section~\ref{sec:photometry}).

The 3I/ATLAS photometry was extracted using a fixed aperture radius (\textit{4 pixels=0.15'}), with the sky background estimated in the same annulus. To reduce short-timescale noise and provide homogeneous sampling for the period analysis, the time series was also analyzed in uniformly binned intervals of and 30 minutes, using inverse-variance weighted averages computed independently for each night.

These ground-based data provide an independent constraint on the rotational modulation of the inner coma brightness and enable direct comparison with the jet-derived period obtained from space-based morphology measurements.\\

\begin{figure*}
\centering
\includegraphics[width=0.95\textwidth]{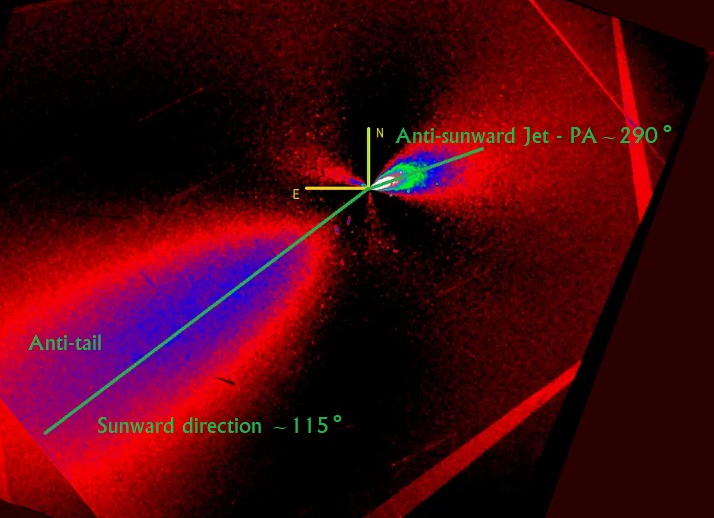}
\caption{Illustration of the anti-sunward jet and the opposing anti-tail in The Larson-Sekanina Rotational Gradient filter of a representative HST image. The position angle (PA) of the anti-sunward jet is marked, using the standard convention N=0° and E=90°.}
\label{fig:residual_phase}
\end{figure*}

\end{document}